\documentclass[onecolumn,10pt,a4paper]{quantumarticle}
\pdfoutput=1
\usepackage[utf8]{inputenc}
\usepackage{comment}
\usepackage{todonotes}
\usepackage{authblk}
\usepackage{amsmath}
\usepackage{amssymb}
\usepackage{graphicx}
\usepackage{subcaption}
\usepackage[font=footnotesize]{caption}
\usepackage[backend=bibtex,style=numeric,sorting=none,citestyle=numeric-comp]{biblatex}

\bibliography{paper}

\begin{document}

\title{Kernel Matrix Completion for Offline Quantum-Enhanced Machine Learning}

\author[1]{Annie Naveh}
\author[1]{Imogen Fitzgerald}
\author[2,3]{Anna Phan}
\author[1]{Andrew Lockwood}
\author[4]{Travis L. Scholten}
\email{Travis.Scholten@ibm.com}
\affil[1]{Woodside Energy Ltd., Perth, WA, Australia}
\affil[2]{IBM Quantum, Melbourne, Victoria, Australia}
\affil[3]{School of Physics, University of Melbourne, Parkville, Victoria, Australia}
\affil[4]{IBM Quantum, IBM T.J. Watson Research Center, Yorktown Heights, New York, USA}

\date{}

\begin{abstract}
Enhancing classical machine learning (ML) algorithms through quantum kernels is a rapidly growing research topic in quantum machine learning (QML). A key challenge in using kernels -- both classical and quantum -- is that ML workflows involve acquiring new observations, for which new kernel values need to be calculated. Transferring data back-and-forth between where the new observations are generated \& a quantum computer incurs a time delay; this delay may exceed the timescales relevant for using the QML algorithm in the first place. In this work, we show quantum kernel matrices can be extended to incorporate new data using a classical (chordal-graph-based) matrix completion algorithm. The minimal sample complexity needed for perfect completion is dependent on matrix rank. We empirically show that (a) quantum kernel matrices can be completed using this algorithm when the minimal sample complexity is met, (b) the error of the completion degrades gracefully in the presence of finite-sampling noise, and (c) the rank of quantum kernel matrices depends weakly on the expressibility of the quantum feature map generating the kernel. Further, on a real-world, industrially-relevant data set, the completion error behaves gracefully even when the minimal sample complexity is not reached.
\end{abstract}


\maketitle

\section{Introduction}
\subsection{Quantum Machine Learning \& Quantum Kernels}
\label{sec:qml-qk}
Quantum machine learning (QML) \cite{biamonte2017quantum,schuld2015introduction,wittek2014quantum} is an interesting and cross-disciplinary research topic which has attracted growing attention in recent years. The cross-pollination of ideas between machine learning (ML) and quantum information has lead to research efforts in four general areas: applying ML techniques into quantum physics experiments (aka, ``ML for Physics") \cite{torlai2020quantum, torlai2017many,scholten2019classifying, ml4physicalsciences}, developing new ML approaches from quantum information ideas, such as tensor networks (aka, ``Quantum-Inspired ML") \cite{menneer1995quantum, stoudenmire2016supervised}, using quantum algorithms to process data (aka, ``Quantum Learning") \cite{lloyd2016quantum,harrow2009quantum,arunachalam2017guest}, and enhancing classical ML algorithms through access to quantum circuits (aka, ``Quantum-Enhanced ML") \cite{havlivcek2019supervised, killoran2019continuous, farhi2018classification, abbas2021power, schuld2021supervised}. 
    
Of these four areas, two -- Quantum Learning and Quantum-Enhanced ML -- involve the use of quantum computers in ML workflows. For near-term, noisy quantum hardware, running high-depth, high-width algorithms will be difficult, and they generally require fault-tolerant hardware. Because of this, the current focus in the research literature is on exploring ``quantum-enhanced" ML algorithms. These algorithms include quantum neural networks \cite{havlivcek2019supervised, killoran2019continuous, farhi2018classification, abbas2021power} and quantum kernels \cite{havlivcek2019supervised, schuld2021supervised}. With these algorithms, a classical ML workflow is \emph{enhanced} by endowing the algorithm with the ability to encode (and process) classical data in quantum state space.
    
In this work, we focus on one family of quantum-enhanced ML algorithms: quantum kernels (also known as the ``quantum kernel estimator" or ``quantum support vector machines"). In ML, a kernel function is a similarity measure between two data points in some vector space of dimension $d$, say $\mathbf{x}_{l}$ and $\mathbf{x}_{m}$, which we denote as $k(\mathbf{x}_{l},\mathbf{x}_{m})$. Kernels are used to implicitly ``lift" the data into a higher-dimensional feature space with dimension $f > d$. Given a kernel function $k$, there exists a \emph{feature map} $\phi(\mathbf{x}): \mathbb{R}^{d} \rightarrow \mathbb{R}^{f}$ such that $k(\mathbf{x}_{l}, \mathbf{x}_{m}) = \langle \phi(\mathbf{x}_{l}), \phi(\mathbf{x}_{m})\rangle$. Here, $\langle \mathbf{a}, \mathbf{b} \rangle$ denotes the inner product between $\mathbf{a}$ and $\mathbf{b}$. Further, note that the kernel function is a symmetric function. The advantage of using the kernel function is that the inner product can be evaluated without having to explicitly map the data into the higher-dimensional feature space. Given a collection of data, the values of the kernel function evaluated on all pairs of data can be arranged as a symmetric and positive-semidefinite matrix $K$, with entries $K_{lm} \equiv k(\mathbf{x}_{l}, \mathbf{x}_{m})$ \cite{hofmann2008}.

Kernels can be used in algorithms for supervised learning, such as classification using support vector classification, or regression using kernel ridge or Gaussian process regression, as well as unsupervised learning, such as clustering using spectral or agglomerative clustering, and dimension reduction using kernel principal component analysis \cite{Goodfellow-et-al-2016}.

Quantum kernels are calculated using quantum circuits. Often, this circuit has an explicit parameterisation in terms of the data to be embedded; circuits of this form go by the name “Parameterised Quantum Circuits” (PQCs). However, there are other circuits \cite{liu2020} which can also be used, but they do not have an explicit parameterisation. A width-$w$ PQC is a unitary operation on $w$ qubits, denoted $U(\boldsymbol{\theta})$, acting on some initial state, usually chosen to be $|0^{\otimes w}\rangle$. The action of $U$ can then be used as a quantum feature map \cite{schuld2019hilbert}, by considering the state
\begin{equation}
\rho(\mathbf{x}) = |\psi(\mathbf{x})\rangle \langle \psi(\mathbf{x})|, \end{equation}
where
\begin{equation}|\psi(\mathbf{x})\rangle = U(\mathbf{x})|0^{\otimes w}\rangle.
\end{equation}
$\rho(\mathbf{x})$ is the ``feature state", the encoding of the data point $\mathbf{x}$ into Hilbert space. The use of the density matrix representation, even for pure states, implies that the feature state does not depend on any global phase which may be associated with the pure state. The corresponding quantum kernel is the overlap between the two feature states:
\begin{equation}
k(\mathbf{x}_{l}, \mathbf{x}_{m}) = \mathrm{Tr}\left(\rho(\mathbf{x}_{l})\rho(\mathbf{x}_{m})\right) = |\langle 0^{\otimes w}|U^{\dagger}(\mathbf{x}_{l})U(\mathbf{x}_{m})|0^{\otimes w}\rangle|^{2}.
\end{equation}
The value of $k(\mathbf{x}_{l}, \mathbf{x}_{m})$ can be estimated by directly implementing the circuit $U(\mathbf{x}_{m})U^{\dagger}(\mathbf{x}_{l})$ on the $|0^{\otimes w}\rangle$ state and then estimating the probability of obtaining the all-zeros bitstring. This estimate will be affected by noise stemming from state preparation and measurement (SPAM) error, gate error, and finite-sampling (``shot") statistics. The values of the quantum kernel function can, like their classical counterparts, be arranged into a matrix and used in any kernel-based ML algorithm.

To date, identifying \emph{particular} PQCs which offer both a \emph{computational} and \emph{practical} advantage for quantum kernels has remained elusive \cite{havlivcek2019supervised}. However, the study of quantum kernels has spawned several fruitful results, including: an unambiguous proof of quantum advantage for a particular problem through the use of quantum kernels \cite{liu2020}, the recognition that geometric considerations play a role for quantum advantage generically \cite{huang2021}, a unification of quantum machine learning models with quantum kernels \cite{schuld2021supervised}, and a demonstration that ``kernel alignment" can help make quantum kernels more useful \cite{glick2021covariant}. These results suggest quantum kernels may be a potentially promising avenue to demonstrate quantum advantage.

Quantum kernels are interesting from a software development perspective, as well. They can be programmed and used by developers through software packages such as Qiskit Machine Learning, TensorFlow Quantum, and PennyLane \cite{Qiskit, broughton2021tensorflow, bergholm2020pennylane}.

\subsection{Extending Kernels to Process Streaming Data}
\label{sec:kernelextensionintro}
Unfortunately, an understanding of how quantum-enhanced ML might be affected by practical considerations of data science workflows has been lacking in the research literature. In particular, existing work has often applied quantum-enhanced ML to canonical data sets such as Iris and MNIST \cite{abbas2021power, hubregtsen2021training,wilson2018quantum, hur2021quantum,mohsen2021image}. To our knowledge, no work has been done studying the implications of quantum-enhanced ML that would process \emph{streaming} data. By streaming data, we mean that new data points are being generated and must be analysed by the algorithm. If a classical ML workflow must process such data, then it must contend with the fact that, over time, its quantum-enhanced variant will need to process those new data points using quantum circuits. Note that this practical consideration would also be applicable even if the entire workflow was being used to try and demonstrate quantum advantage using quantum kernels. That is, the question considered here is essential to understanding how to use quantum kernels in practice.
    
Suppose an original data set of size $N$ has been generated, and the corresponding $N\times N$ quantum kernel matrix $K$ has been computed. Now, suppose $n$ new data points are generated. The new kernel matrix entries to be computed fall into two camps: those \emph{within} the new data set (of which there are $\mathcal{O}(n^{2})$), and those \emph{between} the new data set and the original (of which there are $Nn$); that is, the original $N \times N$ quantum kernel matrix must be \emph{extended} to an $(N+n)\times (N+n)$ one.

The reader may be wondering whether extending the kernel matrix to encompass all of the old and new data is necessary. The answer is ``generally, yes". There do exist specific kernel-based ML algorithms for which full extension would not be necessary. For example, in a support vector machine, the kernel would need to be extended only to encompass kernel values between the new data points and the support vectors. However, for other algorithms -- such as Gaussian process regression -- full extension of the kernel matrix is necessary.

Most QML workflows do not account for the processing of streaming data. In particular, unless the old data resides in storage (indefinitely) near the quantum computer, then naively, both it and all of the new data needs to be transferred from wherever they are stored to the quantum computer. (Note that the data transfer of the new data is an unavoidable cost.) Although the total data transfer in this case scales as $\mathcal{O}(n(N+n))$, and is efficient from a computational complexity standpoint, practical considerations suggest a looming bottleneck. In particular, the overall time cost of this transfer -- along with the time needed to process the data and return estimates of the quantum kernel -- may exceed the timescales which are necessary for using the QML algorithm. The shorter the timescale over which the QML workflow needs to provide actionable insights, the tighter the timing constraints become for processing streaming data in a quantum-enhanced workflow. Although advances in the speed with which circuits can be executed could help by decreasing the time latency \cite{wack2021quality}, the simple fact of the matter is that for streaming data, at some point the naive approach of ``ship all the data, every time" will become time-prohibitive.

However, by examining the \emph{structure} of this problem, we can see that it is naturally framed as a very particular matrix completion problem. It is this framing we turn to next.

\subsection{The Structure of Kernel Matrix Extension}

Extending quantum kernel matrices to encompass new data has a particularly unique structure.  Since $n$ new data points were generated, $\mathcal{O}(n^{2})$ new kernel values need to be calculated at minimum. And by assumption, $\mathcal{O}(N^{2})$ kernel values already exist. Arrange these values into an $(N+n)\times (N+n)$ matrix, with a filled-in $(N\times N)$ upper-left block. Cordon off the $(n\times n)$ lower-right block, and assume those kernel values have been calculated. Assume there are zeros everywhere else.  Clearly, this matrix is \emph{block-diagonal}. (Skip ahead to Figure \ref{fig:u=0} for a visual.) Note that the zero entries in this matrix represent the unknown kernel values between a piece of old \emph{and} new data.

As will be discussed in Section \ref{sec:qks-matrixcompletion}, this arrangement of blocks is an example of what is called a \emph{block-diagonal sparsity pattern}. A more general version of a block-diagonal sparsity pattern can be defined (Section \ref{sec:block-diag}) which incorporates the notion of an \emph{overlap} between the blocks. (Skip ahead to Figure \ref{fig:u>0} for a visual.) When the blocks have overlap, some of the previously-empty kernel values between the old and new data are filled in (meaning they have been calculated using the quantum computer). Note that in this figure, and throughout, the overlap is denoted by $u$. It governs the number of old data points which are sent to the quantum computer.

The still-unknown kernel values can be estimated using matrix completion techniques based on chordal graph algorithms \cite{candes2009exact}. These algorithms achieve perfect accuracy in the ideal case. Crucially, the particular algorithm we consider here does \emph{not} require a guarantee that the extended (that is, $(N+n)\times (N+n)$ kernel matrix) be low rank. By exploiting the structure of kernel matrix extension and these powerful algorithms, we can minimise the data transfer costs, by identifying the minimal value of $u$ necessary.
~\\~\\
The remainder of this paper is organised as follows.  Section \ref{sec:qks-matrixcompletion} formalises the notion of extending a quantum kernel matrix, and presents a description of chordal-graph-based matrix completion. Section \ref{sec:completion-results-uniform}, presents numerical results on kernel matrix extension using a random, synthetic data set and a variety of PQCs. We find completion error goes to zero in a certain regime (namely, when the overlap between the two blocks exceeds the rank of the extended kernel matrix) which reflects best-case performance with respect to the theory of these algorithms. We find that completion error increases in the presence of finite-sampling noise, though the algorithm seems robust to it. Section \ref{sec:kernel-property-prediction} explores the relationship between the \emph{expressibility} of a PQC and the rank of the quantum kernel matrix it can generate. We find that highly-expressible PQCs -- which can more accurately approximate Haar-random circuits -- generate quantum kernel matrices whose rank may scale \emph{exponentially} with circuit width $w$. Although this may suggest that extending quantum kernel matrices is not feasible in practice, Section \ref{sec:woodsidedata} considers extending a quantum kernel matrix based on real-world, Woodside-relevant data; we find that the completion error is significantly lower than for random data when the overlap is not sufficiently large (as compared to the rank). We conclude in Section \ref{sec:conclusions} with some observations about quantum kernel matrix extension.

\section{Formulating Quantum Kernel Matrix Extension as Matrix Completion}
\label{sec:qks-matrixcompletion}
To extend quantum kernel matrices involving unknown entries, we turn to chordal-graph-based matrix completion algorithms to complete the matrix. In this section, we introduce the mathematical formalism necessary to describe these algorithms, as well as the notion of \emph{sparsity patterns}. Readers familiar with this material may proceed to Section \ref{sec:completion-results-uniform}, which presents our numerical results.

\subsection{Kernel Matrix Completion: Problem Formulation}

Matrix completion is the task of completing (filling in) all entries of a matrix given that only some entries are known. An incomplete matrix, which we will denote throughout this work as $K'$, can be viewed as the partial representation of a \emph{full} matrix $K$.

Let $S$ represent the set of \emph{known} matrix elements of $K$ which will be included in $K'$:
\begin{align}
S = \{(l,m) \mid K_{lm}~\text{is known}\}.
\end{align}
Then, the entries of $K'$ in $S$ are the same as $K$, with zeros everywhere else:
\begin{align}
K'_{lm}= \begin{cases}K_{lm}~(l,m) \in S \\ 0~~~~~(l,m)\not\in S
\end{cases}.
\end{align}
Completing $K'$ means estimating the unknown entries $K'_{l'm'} \; \forall \; (l', m') \notin S$, yielding an estimate $\hat{K}$.

Without any restrictions on the completion, estimating the unknown entries of $K'$ is an ill-posed problem, since the unknown entries could be assigned any arbitrary value. Even imposing the conditions that the estimate is symmetric and positive-semidefinite is not enough to guarantee a unique completion.

To overcome this, we take the completion $\hat{K}$ to be the maximum-determinant completion, defined as the solution to the optimisation problem
\begin{align}
\label{eqn:max_det}
\text{maximise} & \; \log\left[\det\left(\hat{K}\right)\right]\\
\label{eqn:max_det_cond_1}
\text{s.t. }& \; \hat{K}_{lm} = K'_{lm} \; \forall \; (l,m) \in S \\
& \hat{K} \geq 0.
\end{align}
This completion minimises the Kullback-Leibler divergence between $\hat{K}$ and $K$ \cite{vandenberghe2015chordal}. It requires the estimate to agree with $K$ on the \emph{known} entries; in general, on the \emph{unknown} entries, $\hat{K}$ and $K$ will differ, unless certain criteria are met.

The success of a matrix completion algorithm strongly depends on $S$ -- intuitively, the larger the set, the greater percentage of the matrix is already specified, leading to more accurate completions. However, perhaps surprisingly, is that not only does the size of the set impact completion accuracy but so does the pattern that it induces with respect to $K'$. This pattern of known vs. unknown matrix entries is called the \textit{sparsity pattern}. In the next section, we introduce the formalisation of sparsity patterns and a particularly important sparsity pattern, the \emph{block-diagonal pattern}.

\subsection{Sparsity Patterns Emerging from  Matrix Extension}
\label{sec:sparse-patterns}
The sparsity patterns studied in this work are those that naturally emerge when using kernel matrices with live streaming data. Recall the formulation from Section \ref{sec:kernelextensionintro}: suppose we have $N$ old data points, and their pairwise kernel values have been calculated and arranged into an $N\times N$ matrix. Then, $n$ new data points are generated. The kernel matrix must now be extended to an $(N+n)\times (N+n)$-sized one.

Suppose that kernel values are calculated for all pairs of the $n$ new data points, giving a filled in $n\times n$ sub-block. The sparsity pattern -- indicating the known kernel values -- is the union of two disjoint sets:
\begin{align}
\label{eqn:S_nooverlap}
S &=  \{1, 2, \dots , N\}^2 \cup \{N + 1, N + 2, \dots , N + n \}^2~\text{and}\\
& \left| \{1, 2, \dots , N\} \cap \{ N + 1, N + 2, \dots, N + n \} \right| = \varnothing.
\end{align}
A pictorial representation of this sparsity pattern is shown in Figure \ref{fig:u=0}. Note that the notation $\{\}^{2}$ means ``take the Cartesian product of the elements of the set".

Notice that the first subset in this union indexes the original $N \times N$ matrix and the second indexes the $n \times n$ sub-matrix corresponding to the new data set; hence, the intersection of these two subsets is the empty set. This means there is no overlap between the two blocks, and has substantial implications for matrix completion. In particular, because kernel values are relative measures, knowing the kernel values $k(\mathbf{x}, \mathbf{y})$ and $k(\mathbf{x}', \mathbf{y}')$ does not provide any indication of what $k(\mathbf{x}, \mathbf{x}')$ might be. This is exactly the problem we run into when calculating kernel values separately for each data set: we do not know any of the kernel values \emph{between} the old and new data points.
Given this, we cannot assume or even expect matrix completion to succeed. 
Phrased another way, \emph{some} amount of the old data needs to be transferred, along with the new.

By transferring some ($u$) old data, the kernel values between old and new data provides information which can be exploited by the matrix completion algorithm. Consider the sparsity pattern 
\begin{align}
\label{eqn:S_someoverlap}
S &= \{1, 2, \dots , N\}^2 \cup \{N - u + 1, N - u + 2, \dots, N + n \}^2~\text{where}\\
&\left| \{1, 2, \dots , N\} \cap \{N - u + 1, N - u + 2, \dots, N + n \} \right| = u.
\end{align}

A pictorial description of this sparsity pattern is presented in Figure \ref{fig:u>0}. Because the cardinality of the intersection of the two sets, $u$, is non-zero, $u$ is called the \emph{overlap}.
Having a non-zero overlap provides information about both the new and old data in a way which fundamentally makes matrix completion possible.

This sparsity pattern is a simple case of what is known as a \emph{block-diagonal} sparsity pattern. In the next subsection, we generalise the ideas presented here, with an eye towards presenting later in Section \ref{sec:matdetcompletion} a chordal-graph-based completion algorithm for filling in matrices with such a sparsity pattern.

\subsection{The Block-diagonal Sparsity Pattern}
\label{sec:block-diag}
\begin{figure}[t]
\centering
\begin{subfigure}[b]{0.49\columnwidth}
\centering
\includegraphics[width=\columnwidth, height =0.65\columnwidth ]{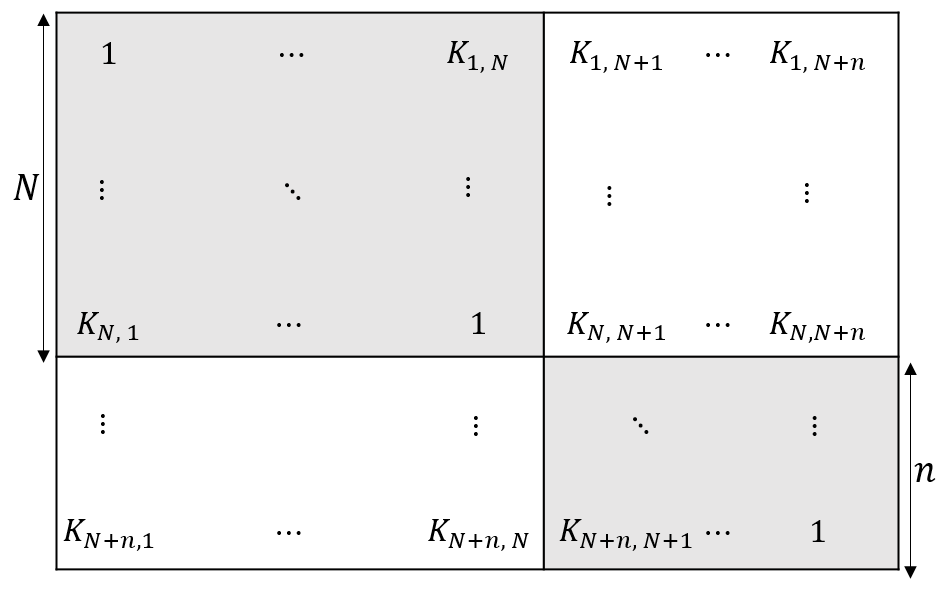}
\vspace{-1.3\baselineskip}
\caption{\textbf{Sparsity pattern indexed by \eqref{eqn:S_nooverlap}: No overlap.}}
\label{fig:u=0} 
\end{subfigure}
\begin{subfigure}[b]{0.49\columnwidth}
\centering
\includegraphics[width=\columnwidth, height = 0.65\columnwidth]{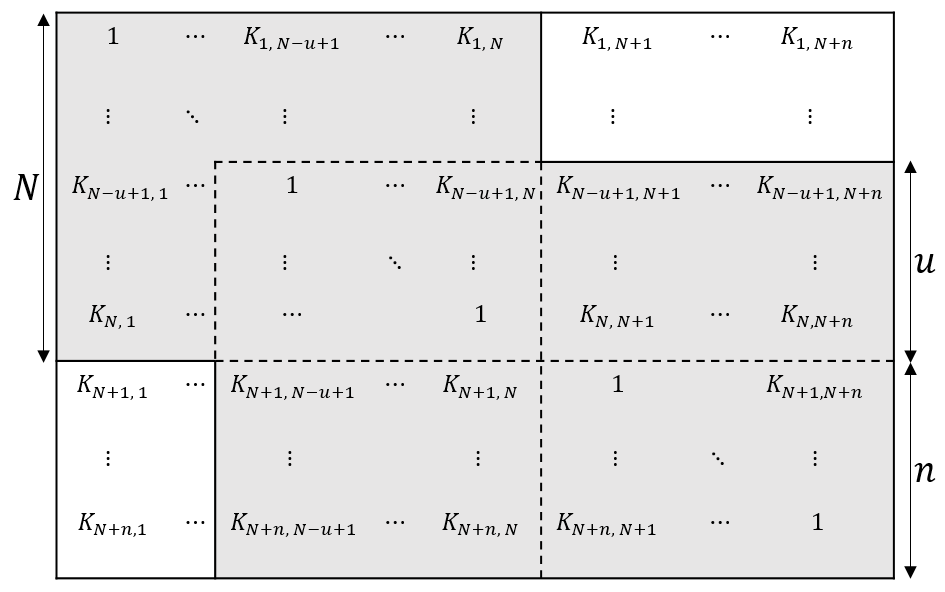}
\vspace{-1.3\baselineskip}
\caption{\textbf{Sparsity pattern indexed by \eqref{eqn:S_someoverlap}: Overlap of $u$.}}
\label{fig:u>0}
\end{subfigure}
\vspace{-0.5\baselineskip}
\caption{\textbf{Block-diagonal sparsity pattern formed from a kernel matrix extension.} An $N \times N$ kernel matrix is extended for $n$ new data points. The resulting $(N + n) \times (N + n)$ extended matrix is sparse if \emph{only some} ($u < N$) of the original data set is re-transferred with the new. In particular the sparsity pattern is block-diagonal with two blocks overlapping each other by $u$. In both figures, the shaded entries of the matrix are those that are known i.e., indexed by $S$ in \eqref{eqn:S_nooverlap} and \eqref{eqn:S_someoverlap} respectively.}
\label{fig:blockdiagonal}
\vspace{-0.7\baselineskip}
\end{figure}
The sparsity patterns in Figure \ref{fig:blockdiagonal} and described in Equations \eqref{eqn:S_nooverlap} and \eqref{eqn:S_someoverlap} belong to a more general family of \textit{block-diagonal} patterns. These more general patterns induce (possibly overlapping) diagonal blocks of \emph{known} entries in $K'$.

Let $\Omega_b$ denote a set of indices for a block (square sub-matrix) along the diagonal. Here, $b$ denotes the $b^{\mathrm{th}}$ block (out of, say, $n_b$ blocks in total). Let $\rho[\Omega_b]$ denote the row indices of $\Omega_b$:
\begin{equation}
    \rho[\Omega_b] = \{l \mid (l,m) \in \Omega_b\} .
\end{equation}
Given this, the number of row indices for the intersection of two subsequent blocks $\Omega_b \cap \Omega_{b+1}$ is their overlap $u$:
\begin{align}
\vert \rho[\Omega_{b} \cap \Omega_{b+1}] \vert = u.
\end{align}
A sparsity pattern can be defined with respect to these blocks as their union:
\begin{equation}
\label{eq:blockdiagonalsparsity}
S = \bigcup\limits_{b=1}^{n_b}\Omega_b.
\end{equation}
Using this notation, sets \eqref{eqn:S_nooverlap} and \eqref{eqn:S_someoverlap} described earlier can be seen to induce a block-diagonal sparsity pattern with two blocks $\Omega_1$ and $\Omega_2$. The first block $\Omega_1$ overlaps the consecutive block $\Omega_2$ by an amount determined by the original data re-transferred i.e., $ u = 0$ and $0 < u \le \vert \rho[\Omega_1] \vert $ for sets \eqref{eqn:S_nooverlap} and \eqref{eqn:S_someoverlap} respectively.

\begin{figure}[t]
\centering
\includegraphics[width=0.5\columnwidth, height = 0.3\columnwidth]{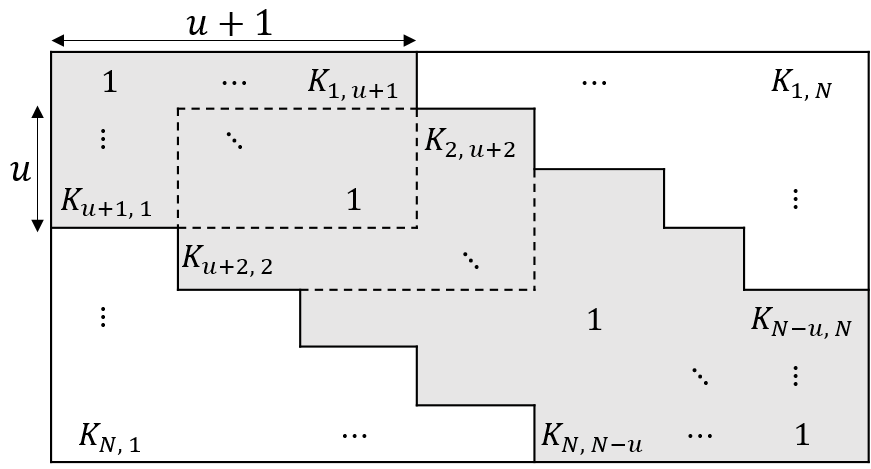} \caption{\textbf{Band sparsity pattern with bandwidth $w = u$.} The band sparsity pattern is the limiting case of the block-diagonal for minimum sampling complexity and maximum overlap. The shaded entries of the matrix represent the known values.}
\label{fig:band}
\vspace{-0.7\baselineskip}
\end{figure}

There is a special limiting case of the generalised block-diagonal in which the overlap between all blocks is maximised, inducing a \emph{band} pattern. (This is depicted visually in Figure \ref{fig:band}.) In this pattern all blocks are the same size, $\vert \rho[\Omega_k] \vert = w + 1$, for bandwidth parameter $w$ such that $u = w$. The band pattern is noteworthy because it represents the minimum sampling complexity (fewest calls to the quantum computer) in the family of block-diagonal patterns.

We focus our work on block-diagonal sparsity patterns mainly because, as discussed in Section \ref{sec:sparse-patterns}, these patterns naturally emerge during matrix extension. What's more, given these sparsity patterns, there are efficient graph-based algorithms for doing the completion. It is to this topic we turn next, to show how block-diagonal sparsity patterns can be understood as \emph{chordal} graphs and to introduce a concomitant matrix completion algorithm.

\subsection{Block-diagonal Sparse Matrices as Chordal Graphs}
\label{sec:chordal-graph-approach}
The graph representation of a symmetric sparsity pattern $S$ is called its \textit{sparsity graph}, $G = (V,E)$. Here, the vertices denote the column (or row) indices of the matrix, and $|V|$ is its size. $G$ contains \emph{undirected} edges such that vertices $l$ and $m$ are connected by an edge if, and only if,  $(l,m) \in S$. 

Block-diagonal sparsity patterns induce corresponding sparsity graphs that are \textit{chordal}: every cycle of length greater than three has a chord. (A chord is an edge that is not part of the cycle but connects two vertices of the cycle.) Although not immediately obvious, it is exactly this property which lends block-diagonal patterns so well to matrix completion.

More generally, chordality is a fundamental property in sparse matrix theory and it is known that exploiting the chordal structure of a sparse matrix can drastically reduce the problem size in semidefinite programming (including positive-semidefinite matrix completion) \cite{grone1984positive, fukuda2001exploiting, nakata2003exploiting}. Of particular note is reference \cite{andersen2010implementation} which shows that for a band sparsity pattern the time per iteration scales at least quadratically with matrix size for nonchordal solvers, whereas, for the chordal solvers it scales only linearly.

Beyond this practical consideration of the time efficiency of solvers which exploit chordal structure, another, more fundamental consideration highlights why the chordal graph structure is important.

It is known that when $K'$ has a chordal sparsity pattern, the inverse of the maximum-determinant completion $\hat{K}$ (that is, the solution to Equation \eqref{eqn:max_det}) has the \emph{same} sparsity pattern as $K'$ \cite{andersen2010support}. That is, if $S$ is the sparsity pattern of $K'$, then the maximum-determinant completion $\hat{K}$ satisfies
\begin{align}
(\hat{K}^{-1})_{l'm'} = 0 \; \forall \; (l', m') \notin S.
\label{eqn:K_inv}
\end{align}
If $\hat{K} > 0$, its inverse is also positive, meaning it has a Cholesky decomposition:
\begin{align}
\hat{K}^{-1} = LDL^{T},
\label{eqn:cholesky_decomp}
\end{align}
where $L$ is a unit lower triangle and $D$ is a positive diagonal. 
Note that if $\hat{K} \geq 0$ (that is, it does have zero eigenvalues), then a Cholesky decomposition can be performed using its \emph{pseudoinverse} $\hat{K}^{+}$.
Combining Equations \eqref{eqn:K_inv} and \eqref{eqn:cholesky_decomp}, it follows that
\begin{align} 
(L+L^{T})_{l'm'} = 0 \; \forall \; (l', m') \notin S.
\label{eqn:L_sparse_pattern}
\end{align}
Equation \eqref{eqn:L_sparse_pattern} is perhaps the most powerful property of chordal graphs and underpins most chordal-based algorithms \cite{andersen2010implementation}. Indeed it is this relationship that lets us efficiently find the maximum-determinant completion for incomplete matrices with block-diagonal patterns. The reason Equation \eqref{eqn:L_sparse_pattern} is so powerful is that it provides an important constraint for an \emph{iterative} chordal solver, described in detail in \cite{vandenberghe2015chordal}. We provide a high-level description of this solver in the next section.

\subsection{Overview of the Maximum-determinant Completion Algorithm}
\label{sec:matdetcompletion}
In this section, we briefly describe a maximum-determinant completion algorithm for incomplete matrices whose sparsity pattern has a chordal graph structure. For more details, see reference \cite{vandenberghe2015chordal}.

\begin{figure}[t]
\centering
\begin{subfigure}[b]{0.32\columnwidth}
\centering
\includegraphics[width=0.7\columnwidth, height =0.7\columnwidth ]{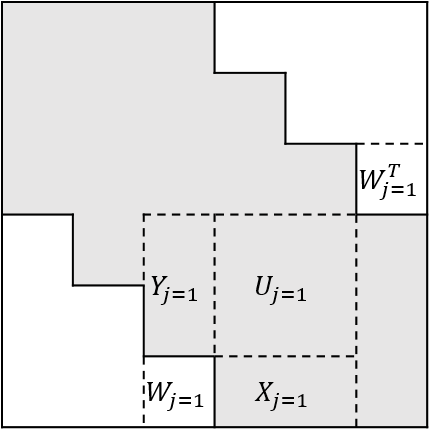}
\caption{\textbf{First iteration} ($j = 1$).}
\label{fig:max_det_alg_iter_1} 
\end{subfigure}
\begin{subfigure}[b]{0.32\columnwidth}
\centering
\includegraphics[width=0.7\columnwidth, height =0.7\columnwidth ]{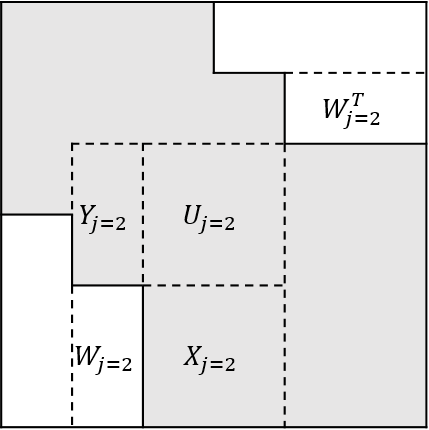}
\caption{\textbf{Second iteration} ($j = 2$).}
\label{fig:max_det_alg_iter_2} 
\end{subfigure}
\vspace{-0.5\baselineskip}
\centering
\begin{subfigure}[b]{0.32\columnwidth}
\centering
\includegraphics[width=0.7\columnwidth, height =0.7\columnwidth ]{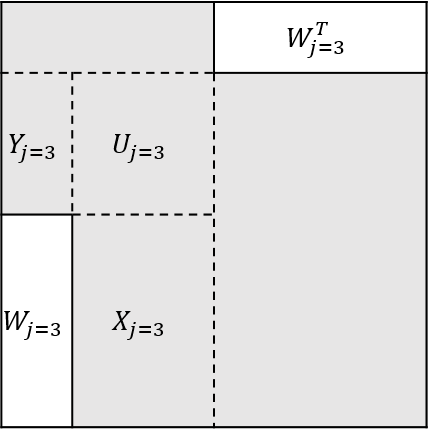}
\caption{\textbf{Third iteration} ($j = 3$).}
\label{fig:max_det_alg_iter_3} 
\end{subfigure}
\caption{\textbf{First three iterations of the maximum-determinant algorithm} as described in reference \cite{vandenberghe2015chordal}. The algorithm returns a positive-semidefinite completion of a block-diagonal sparse matrix by iterating over the supernodes, starting at the bottom right corner. At each step $j$ the surrounding matrix entries at $W_j$ are completed (using known matrix entries indexed by $X_j$, $Y_j$ and $U_j$) by solving Equation \eqref{eqn:max_det_j_pseudo}. }
\label{fig:max_det_alg_iters} 
\vspace{-0.7\baselineskip}
\end{figure}

Given an incomplete matrix $K'$ with block-diagonal sparsity pattern $S$, the maximum-determinant algorithm estimates $\hat{K}$ by iterating over the indices of the sparsity pattern. Recall that for matrices with a block-diagonal sparsity pattern, we can represent $S$ as in Equation \eqref{eq:blockdiagonalsparsity}. The algorithm iterates over the \emph{supernodes} of each block, which are the block's unique row indices. Letting $\text{snd}(j)$ denote the supernode considered at step $j$ of the algorithm, it is given by
\begin{align}
\text{snd}(j) = \rho[\Omega_j] \backslash \rho[\Omega_j \cap \Omega_{j+1}],
\end{align}
where $\Omega_j \cap \Omega_{j+1}$ indexes the $u \times u$ overlap sub-block which we will denote by 
\begin{align}
U_j = \Omega_j \cap \Omega_{j+1}. 
\end{align}
These supernodes are ordered in reverse order so here $\Omega_1$ refers to the bottom-right block and $\Omega_2$ the next block up the diagonal etc.
As demonstrated in Figure \ref{fig:max_det_alg_iters}, additional sub-blocks, at each step $j$, are defined as
\begin{align}
X_j &= \{(l,m) \mid l \in \text{snd}(j), m \in \rho[U_j]\} \\ 
Y_j &=  \{(l,m) \mid l \in \rho[U_j], m \in \text{snd}(j+1) \} \\
W_j &= \{(l,m) \mid l \in \text{snd}(j), m \in \text{snd}(j+1)\}.
\end{align}
At each iteration the surrounding sparse entries indexed by the $W_j$ are filled in by solving Equation \eqref{eqn:cholesky_decomp} subject to \eqref{eqn:L_sparse_pattern} such that,
\begin{align}
    \hat{K}_{W_j} = K_{X_j} K_{U_j}^{+} K_{Y_j}.
    \label{eqn:max_det_j_pseudo}
\end{align}

By performing a walk over the \emph{supernodal elimination tree} -- a particular ordering of the supernodes -- this completion algorithm can fill in the unknown entries of $K'$.

Recall Section \ref{sec:block-diag} raised the point that if matrix completion was to have any chance to succeed, some kernel values between adjacent blocks must be known; that is, the blocks need some overlap. This intuition is solidified by noticing that Equation \eqref{eqn:max_det_j_pseudo} depends on the $u \times u$ overlap sub-block $U_j$. Clearly $U_j$ is essential for completion, but we cannot expect that for any $u > 0$ , $U_j$ will provide the necessary information about the kernel values between the blocks. In fact, this observation is only useful if we have some metric to tell us exactly how large $u$ must be for successful matrix completion. It is known that for the recovery of a symmetric positive-semidefinite matrix by \emph{any} method, a \emph{necessary} condition for successful completion is:
\begin{align}
\frac{u}{r} \ge 1,
\label{eq:bishop_bounds}
\end{align}
with $r$ the rank of $K$ \cite{bishop2014deterministic}. We can see this by considering that the rank is a measure of the amount of unique information in the matrix, hence we need at least this much redundant information between the blocks to fill in unknown entries.
Because Equation \eqref{eq:bishop_bounds} is only a \emph{necessary}, but not \emph{sufficient} condition, there are no guarantees that completion will succeed whenever $u/r \ge 1$.

To test whether matrix completion is possible for quantum kernel matrices with a block-diagonal pattern, we set up a numerical study, which is detailed in the next section. 

\section{Numerical Results: Extending Quantum Kernel Matrices}
\label{sec:completion-results-uniform}

Section \ref{sec:qks-matrixcompletion} introduced the mathematical machinery necessary to understand extending quantum kernel matrices with a block-diagonal sparsity pattern (Section \ref{sec:block-diag}) using a chordal-graph-based solver (Section \ref{sec:chordal-graph-approach}) for producing maximum-determinant completions. In this section, we describe the numerical experiments we performed. Our results are summarised in Figures \ref{fig:band completion} and \ref{fig:block completion}.

\subsection{Problem Set-up}
\label{sec:prob-setup}

Given a PQC $U(\boldsymbol{\theta})$ and a kernel matrix size $N$, we numerically simulate each kernel matrix entry using simulators available through Qiskit \cite{Qiskit}. This matrix serves as the ground truth $K$, which is then subsampled to form sparse matrices $K'$ with a block-diagonal sparsity pattern, $S$. The maximum-determinant, positive-semidefinite completion of $K'$, $\hat{K}$, is then found, and the accuracy of completion is quantified as the difference between $\hat{K}$ and $K$.

\subsubsection{Ground Truth Quantum Kernel Matrices}
\label{subsubsec:ground-truth}
We use 19 different PQCs as defined in \cite{sim2019expressibility}. These circuits are built into quantum feature maps by varying the circuit width (number of qubits) and depth (by changing the number of \emph{layers}, or repetitions, of the template). Each circuit build realisation is used to generate a quantum kernel matrix from a synthetic, uniformly distributed, data set. We present results for $N =500$ in this section.

We generate kernel matrices for both the idealised (noise-free) case, as well as varying levels of finite-sampling (``shot") noise. 

It should be emphasised that the ground truth quantum kernel matrices are built from i.i.d (independent and identically distributed) random data from a uniform distribution; which is not wholly reflective of the types of structured data sets that would be used in real-world ML applications. This uniformity is useful for the purpose of the analysis (avoiding any bias from the specific correlations of a given data set); however, understanding the impact of a structured data set on completion is an important consideration for industrially-relevant workflows. We explore these ideas in Section \ref{sec:woodsidedata}.

\subsubsection{Block-diagonal Subsampling}
The ground truth quantum kernel matrices are each subsampled in a band sparsity pattern with varying bandwidth parameter (see Figure \ref{fig:band}). As discussed in Section \ref{sec:block-diag}, the band pattern is the limiting case of the block-diagonal in which the sampling complexity is minimised. Incorporating a band pattern into our study, therefore, quantifies the quantum computations necessary for successful completion in the `best-case'. 

Additionally to a band pattern, the ground truth matrices are subsampled to mimic the extension of a $450 \times 450$ matrix to include $50$ new items; that is, a block-diagonal pattern with two blocks. The resultant $500 \times 500$ sparse matrix has an upper block of size $450 \times 450$ (the original matrix), and lower block with size that is varied from $50 \times 50$ to $500 \times 500$. The purpose of varying the size of the second block is so that the overlap $u$ between them is varied in the range $ 0 \le u \le 450$. 

\subsubsection{Reconstruction via Matrix Completion}

The maximum-determinant positive-semidefinite completion, $\hat{K}$, of each sparse matrix is found using the chordal matrix implementation in the chompack software package \cite{dahl2009chompack}. The completion error is quantified as 
\begin{align}
\text{Error} = \frac{\|K_{\bar{S}} - \hat{K}_{\bar{S}} \|_F}{\|K_{\bar{S}}\|_F},
\label{eq:error}
\end{align}
where $\|.\|_F$ is the Frobenius norm, and $\bar{S}$ is the complement of $S$ with respect to the indices of $K$. Recall that, per Equation \eqref{eqn:max_det_cond_1}, $\hat{K}$ and $K$ will agree on the known elements of $K$ anyway; this definition of error removes the effect of the known elements in the denominator.

The error satisfies the bounds $0 \leq \mathrm{Error} \leq 1$. The lower bound follows trivially from the fact that the numerator and denominator are matrix norms. The upper bound follows from the fact that as $u \rightarrow 0$, Equation \eqref{eqn:L_sparse_pattern} necessitates that the elements of $\hat{K}_{\bar{S}}$ approach 0.

There is a subtle complication when completing noisy quantum kernel matrices as the noise may corrupt the positive-semidefinite property of $K$ and hence a positive-semidefinite completion of the corresponding sparse matrix may not exist. This is remediated by an important observation that a sparse matrix with a block-diagonal sparsity pattern has a positive-semidefinite completion if and only if each block in the pattern is itself positive-semidefinite \cite{bishop2014deterministic, vandenberghe2015chordal}. Therefore, for the sparse matrices with block-diagonal elements generated with noise, each noisy block is replaced with the nearest correlation matrix that is positive-semidefinite.

\subsection{Completion Results}

\begin{figure}[t]
\centering
\begin{subfigure}[b]{0.49\columnwidth}
\centering
\includegraphics[width=\columnwidth, height = 0.6\columnwidth]{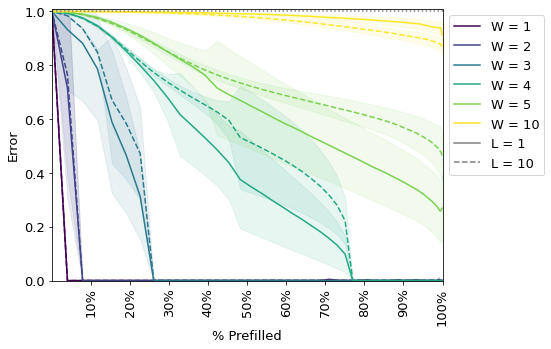}
\caption{\textbf{Noise-free case.}}
\label{fig:band noisefree} 
\end{subfigure}
\begin{subfigure}[b]{0.49\columnwidth}
\centering
\includegraphics[width=\columnwidth, height = 0.6\columnwidth]{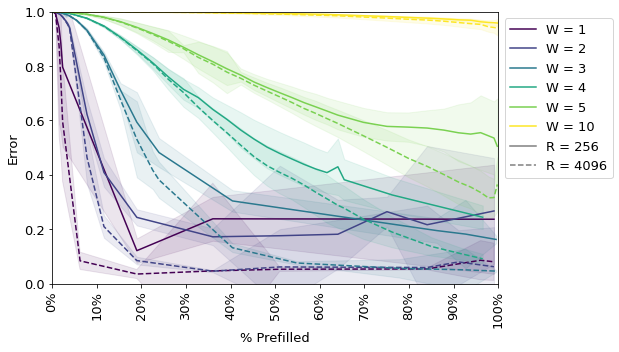}
\caption{\textbf{Noisy (``shot" noise) case.}}
\label{fig:band noisy} 
\end{subfigure}

\caption{\textbf{Completing quantum kernel matrices with a band sparsity pattern using matrix completion.}\\W is the width of the PQC, and L the number of repetitions of its base template. The x-axis captures the percentage of the matrix pre-populated before completion and, as the matrices are symmetric, is calculated over the lower-diagonal (excluding the main diagonal).\\\textbf{Left:} Quantum kernel matrices (generated without any noise) can be completed with full accuracy from only a sample of given entries. \textbf{Right:} The completion error for quantum kernel matrices generated with finite-sampling (``shot") noise (R) plateaus for greater sampling complexity, here L=1.\\
In both figures, the median across the 19 PQCs is calculated and one standard deviation above and below is shaded.}
\label{fig:band completion}

\end{figure}

\begin{figure}[t]
\centering
\begin{subfigure}[b]{0.45\columnwidth}
\centering
\includegraphics[width=\columnwidth, height = 0.7\columnwidth]{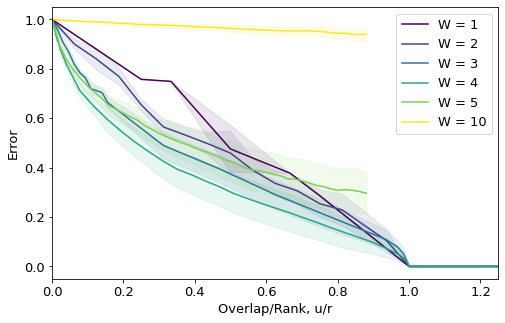}
\caption{\textbf{Noise-free case.}}
\label{fig:block noisefree} 
\end{subfigure}
\begin{subfigure}[b]{0.45\columnwidth}
\centering
\includegraphics[width=\columnwidth, height = 0.7\columnwidth]{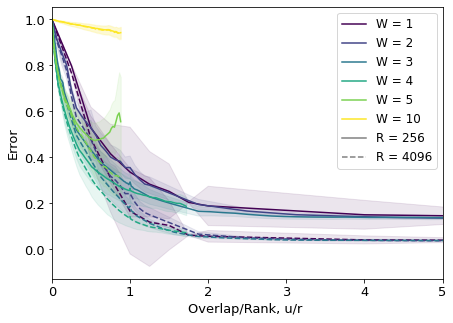}
\caption{\textbf{Noisy (``shot" noise) case.}}
\label{fig:block noisy} 
\end{subfigure}

\caption{\textbf{Completing quantum kernel matrices with a block-diagonal pattern using matrix completion.}\\In both plots, W is the width of the PQC and L is fixed to 1; the median across the 19 PQCs is calculated and one standard deviation above and below is shaded.\\\textbf{Left:} Completion error goes to zero once the overlap between the blocks $u$ exceeds $r$, the rank of the ground truth matrix. \textbf{Right:} Completion accuracy degrades gracefully in the presence of finite-sampling (``shot") noise (R).}
\label{fig:block completion}
\end{figure}

Figure \ref{fig:band noisefree} presents completion results for the \emph{band} sparsity pattern. It shows that noise-free quantum kernel matrices can be recovered with full accuracy from only a set of known elements. As the bandwidth of the band sparsity pattern is increased, the sampling complexity -- measured as the percentage of the matrix pre-populated -- also increases resulting in a higher accuracy of matrix completion. At some critical sampling complexity, which depends highly on the width of the circuit, the completion error falls to zero indicating that the ground truth matrix can be reconstructed with full accuracy. The lower the circuit width, and (in general) the fewer repetitions of the base template $L$, the lower the completion error for a given sampling complexity. That is, kernel matrices generated from higher-width PQCs require more of the matrix to be filled in beforehand for the same completion accuracy. 

The error for the noisy matrices, see Figure \ref{fig:band noisy}, also decreases for greater bandwidth but does not fall all the way to zero, instead plateauing. Its encouraging to see that even with quite a lot of shot-noise ($R = 256$) the maximum-determinant algorithm remains robust and completion error degrades gracefully in the presence of this noise.

Figure \ref{fig:block completion} presents completion results for the \emph{block-diagonal} sparsity pattern with two blocks. We see that when $u/r\geq 1$, completion error goes to zero (for noise-free kernel values). This suggests that the \emph{necessary} condition for matrix completion to succeed (Equation \eqref{eq:bishop_bounds}) is \emph{sufficient} in this case. For noisy data, the completion error degrades gracefully as well. Further, the error introduced from the shot-noise appears to be independent of circuit width, and decreases as more shots are taken.

For extending quantum kernel matrices using a block-diagonal sparsity pattern, Equation \eqref{eq:bishop_bounds} tells us that the relationship between the block \emph{overlap} -- which is in our control, and represents the number of old data points which must be transferred -- and the \emph{rank} of the extended matrix is key to whether perfect completion is even possible in principle. Unfortunately, there is a subtle problem with this: knowing $r$ requires knowledge of $K$, which would not known \emph{a priori} in a practical scenario. Indeed if the full target matrix $K$ was known there would be no need for matrix completion techniques in the first place! This dilemma can be circumvented if $r$ is not \emph{calculated} directly but instead \emph{predicted} by some other means not requiring knowledge of the full $K$. It is to this problem which we turn next.

\section{Predicting Quantum Kernel Matrix Properties from PQC Properties}
\label{sec:kernel-property-prediction}
The importance of the rank of quantum kernel matrices arose in the previous section because it sets a lower bound on the necessary overlap between the blocks to ensure perfect completion is possible (on ideal, noise-free data). But, knowing this rank in advance would be difficult, in part because the new data for which the kernel matrix is being extended is not generally known. What \emph{is} known is the PQC being used. In this section, we show there is a relationship between a recently-introduced property of PQCs -- their \emph{expressibility} \cite{sim2019expressibility} -- and the rank of the quantum kernel matrices they can generate.

There are two simple upper bounds on the rank of an $N\times N$ quantum kernel matrix. The first is $N$ itself. Another upper bound can be derived by observing that an $w$-qubit state $\rho$ can be viewed as a vector in a $4^{w}$-dimensional real vector space (e.g., by expanding $\rho$ in a Pauli basis). Then, the kernel matrix entries are given by
\begin{equation}
    K_{lm} = \mathrm{Tr}(\rho_{j}\rho_{k}) = \mathrm{Tr}\left(\sum_{p=1}^{4^{w}}c_{jp}\sigma_{q}\sum_{q=1}^{4^{w}}c_{qk}\sigma_{q}\right) = \sum_{p,q=1}^{4^{w}}c_{jp}c_{kq}\underbrace{\mathrm{Tr}(\sigma_{p}\sigma_{q})}_{\propto\delta_{pq}} \propto \sum_{p=1}^{4^{w}}c_{jp}c_{kp} \propto \mathbf{c}_{j}\cdot \mathbf{c}_{k},
\end{equation}
where $\mathbf{c}_{\alpha}$ is the generalised Bloch vector for $\rho_{\alpha}$. There can be at most $4^{w}$ such linearly independent vectors (because that is the dimension of the vector space containing $\mathbf{c}_{\alpha}$). As the quantum kernel matrix is also a Gram matrix of such vectors, another upper bound on the rank of $K$ is $4^{w}$.

\begin{figure}
\centering
\begin{subfigure}[t]{.45\columnwidth}
\centering
\includegraphics[width=\columnwidth]{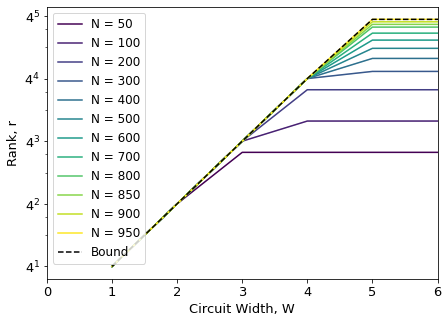}
\caption{\textbf{Haar-random circuits.}}
\label{fig:haar-rank} 
\end{subfigure}
\hspace{2pt}
\begin{subfigure}[t]{0.45\columnwidth}
\centering
\includegraphics[width=\columnwidth]{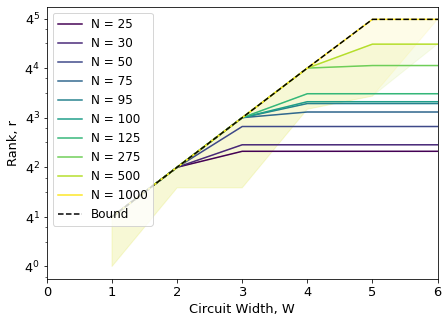}
\caption{\textbf{Parameterised quantum circuits (PQCs).}}
\label{fig:pqc-rank} 
\end{subfigure}
\caption{\textbf{Relationship between circuit width and rank of the quantum kernel matrices generated.}\\ The dashed line indicates the bound on matrix rank from Equation \eqref{eq:rank-bound} (for $N=950$ and $N=1000$ respectively).\\\textbf{Left}: Haar-random circuits appear to saturate the rank bound. Here the data \emph{is not aggregated}, and consists of 10 realisations of Haar-random circuits for each matrix size N. \textbf{Right}: PQCs do not appear to saturate the rank bound, but do display behaviour similar to Haar-random circuits in the regime when $N > 4^{w}.$ The data here includes realisations of PQCs from \cite{sim2019expressibility} with a number of layers $L > 1$. The median across all realisations is plotted as a line whilst the shaded region captures the full range of results.}
\label{fig:ranks}
\end{figure}

Therefore, the rank of an $N\times N$ quantum kernel matrix $K$ generated through the action of a width-$w$ quantum circuit satisfies
\begin{equation}
\label{eq:rank-bound}
    \mathrm{rank}(K) \leq \min(N, 4^{w}).
\end{equation}
When $N > 4^{w}$ then $\mathrm{rank}(K)$ isn't limited by the size of the kernel matrix. It is this regime which we look at in more detail. Numerical evidence (Figure \ref{fig:haar-rank}) suggests that for circuits which are Haar-random on $w$ qubits, the $4^{w}$ upper bound is \emph{saturated}. That is, if $N > 4^{w}$, the rank of the kernel matrix is $4^{w}$. In contrast, numerical evidence from the 19 PQCs introduced in Section \ref{subsubsec:ground-truth} suggests that, in general, PQCs \emph{do not} saturate this bound (Figure \ref{fig:pqc-rank}).

However, the behaviour of the PQCs seems very similar to Haar-random circuits. This leads to the intuition that if a PQC could accurately approximate a Haar-random circuit, then the rank of the quantum kernel matrices it can generate should be similar, if not the same, as the rank of quantum kernel matrices generated by Haar-random circuits.

We sharpen this intuition by using a recently-introduced measure of how well as PQC can approximate a Haar-random circuit, called the \emph{expressibility} \cite{sim2019expressibility}. Given a $w$-qubit PQC $U(\boldsymbol{\theta})$, consider the distribution of fidelities
\begin{equation}
    F_{jk} = |\langle 0^{\otimes w}|U^{\dagger}(\boldsymbol{\theta}_{j})U(\boldsymbol{\theta}_{k})|0^{\otimes w}\rangle|^{2}.
\end{equation}
Let $\mathrm{Pr}_{\mathrm{PQC}}(F)$ denote the emprical distribution of these fidelity values. And let $\mathrm{Pr}_{\mathrm{Haar}}(F)$ denote the distribution of pairwise fidelities under Haar-random unitaries; this distribution is known analytically, and is given by
\begin{equation}
    \mathrm{Pr}_{\mathrm{Haar}}(F) = (2^{w}-1)(1-F)^{2^{w}-2}.
\end{equation}
The expressibility of $U$ is given by the Kullback-Leibler divergence between these two distributions:
\begin{equation}
    e(U) = \mathrm{KL}(\mathrm{Pr}_{\mathrm{PQC}}(F) || \mathrm{Pr}_{\mathrm{Haar}}(F)).
\end{equation}
While the Kullback-Leibler divergence is not a proper distance measure on probability distributions, it has the property that it is zero if, and only if, the two distributions are the same. For this reason, we say a circuit has high expressibility if $e(U)$ is close to zero. In what follows, we will work with $-\log(e(U))$ instead, to facilitate our analysis.

\begin{figure}
    \centering
    \includegraphics[width=\columnwidth]{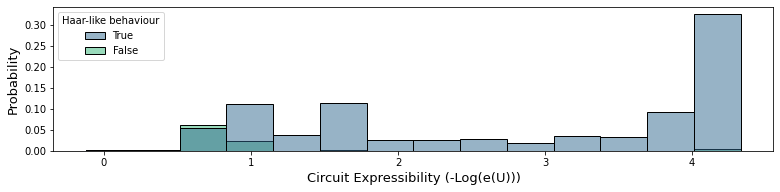}
    \caption{\textbf{Relationship between circuit expressibility and scaling of quantum kernel matrix rank.}\\ Circuit expressibility, introduced in \cite{sim2019expressibility}, is a quantification of how well a PQC approximates a Haar-random circuit. Here, we take the x-axis to be $-\log(e(U))$, meaning higher values of $x$ correspond to more-expressible circuits. The bars are coloured with respect to whether (given a PQC, matrix size and circuit width) the rank of the quantum kernel matrix generated by that PQC is the same as that generated by a Haar-random circuit. (Note: for all data here, $N > 4^{w}$). For PQCs with a lower expressibility, the kernel matrices they generate do not follow the behaviour of Haar-random circuits, whereas for PQCs with higher expressibility, they do.}
    \label{fig:pqc-expr-scaling-2}
\end{figure}

We consider the regime where $N> 4^{w}$, as we know that the rank of kernel matrices generated by Haar-random circuits will be $4^{w}$ in that case. We also know that the rank of kernel matrices generated by PQCs cannot be \emph{higher} than $4^{w}$. Figure \ref{fig:pqc-expr-scaling-2} shows a histogram of $-\log(e(U))$ for a variety of circuits, where a circuit is labeled as ``True" if the rank of the quantum kernel matrix it generates is the same as that for a Haar-random matrix, and ``False" otherwise.

For circuits with higher expressibility (large $-\log(e(U))$), we see that they generally have the same behaviour as Haar-random circuits, whereas when the circuit has low expressibility (low $-\log(e(U))$), then it is not always the case that the behaviour is the same. That is, more-expressible PQCs will generally have the same scaling on the rank of the quantum kernel matrices they generate as Haar-random circuits.

The numerics presented thus far have used \emph{random} choices for the parameters $\boldsymbol{\theta}$. However, real-world data \emph{is not} random, and often exhibits correlations. In the next section, we study the impact of structured data on our analysis, by studying an oil and gas data set from Woodside Energy.

\section{Application: Woodside-Relevant Data Set}
\label{sec:woodsidedata}
Where the data comprise measurements of the state, properties and actions of a physical system, we expect the relationship between features to be correlated, in contrast to the previously considered case where the PQC parameters $\boldsymbol{\theta}$ were i.i.d uniformly random. In many practical modelling problems, despite the physics governing these interactions being well defined, it is impossible to calculate these high dimensional non-linear relationships in a useful time frame.  The current era of ``digital transformation" has seen the promise realised of ML techniques performing high dimensional regression on non-linear problems, given suitably acquired training data. In some cases, they provide a rapid and sufficiently accurate alternative to numerical modelling of the physics of the system. The potential of a quantum advantage motivates integrating a quantum computer into this workflow.

To better capture this type of operational QML workflow, we study the impact of processing such ``real-world" data using the completion approach outlined in Section \ref{sec:qks-matrixcompletion}.  This data set consists of 13 feature variables, containing information about operating conditions and the internal state of a gas liquefaction plant, known as a ``train". These variables are a mixture of controlled variables, such as the quantity, pressure and composition of the feed gas and refrigerants used in the cryogenic heat exchangers. Others are environmental and uncontrolled, such as ambient temperature and pressure.  Regression variables generated using the kernel method may describe the performance of the train; for example, the quantity of produced liquefied gas, or the efficiency of the plant measured as a ratio of enthalpy changes in the gas to the power consumed. 
Despite the dynamic nature of the liquefaction train, it is possible to predict the output variables using standard open source ML methods such as multilinear regression, dynamic linear modeling and various network based approaches. These are able, with minimal hypertuning, to predict the rate at which liquefied product will be produced to $>95\%$ accuracy over a 24 hour time window, providing that a precise weather forecast is available and under the assumption that any control adjustments are within normal operating procedure.

Due to the structure within this data set, we may expect that the corresponding quantum kernel matrices generated will have different properties to the matrices produced from the i.i.d random data sets. For example, the ambient temperature is found to be inversely correlated with production, as it adversely affects both the refrigeration and power generation within the train. This inverse relationship is complex across the high range of temperatures experienced at the site, but appears linear within temperature ranges defining discrete modes of operation. These distinct modes are caused by different rate-limiting constraints on the plant that change with ambient temperature, such as volume of inflow, ability to clean the gas feed, the rate at which the heat exchangers can chill the refrigerant, and the amount of power available to operate the compressors in the refrigerant system. These types of complex dynamics are captured as structure in the data set. For these reasons, we expect the corresponding quantum kernel matrices will behave differently from those generated using a synthetic, random data set.

We replicate the analysis presented in Section \ref{sec:completion-results-uniform} with the same PQCs, kernel matrix sizes, and completion algorithm. One point to note: for data embedding in a quantum feature map, we require that the number of PQC parameters must match the number of features of the data set. To achieve this, the number of features are truncated to match the number of PQC parameters for each circuit build. Circuits for which the number of parameters exceed 13 (the number of features in the data set) are excluded from the analysis. This is the case in PQC IDs 5 to 8, which have a very high ratio of parameters to circuit width.

Ground truth kernel matrices of size $500 \times 500$ are generated and then subsampled in a block-diagonal pattern with two blocks and varying amount of overlap, $u$, as described in Section \ref{sec:prob-setup}. The matrix completion methods of Section \ref{sec:qks-matrixcompletion} are used to reconstruct the full matrix on a noise-free simulator and this estimate is compared to the ground truth. For consistency, the error of completion is again calculated using Equation \eqref{eq:error}. 

\begin{figure}[h]
\centering
\begin{subfigure}[t]{0.45\columnwidth}
\centering
\includegraphics[width=\columnwidth]{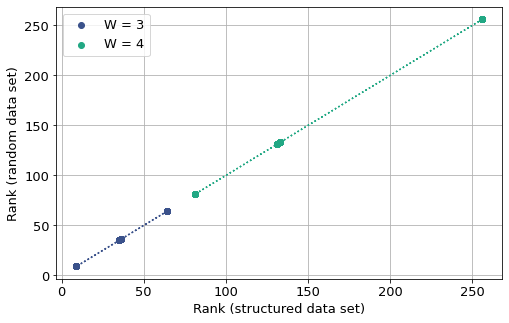}
\caption{\textbf{Comparing matrix rank.}}
\label{fig:techmax-rank} 
\end{subfigure}
\begin{subfigure}[t]{0.45\columnwidth}
\centering
\includegraphics[width=\columnwidth]{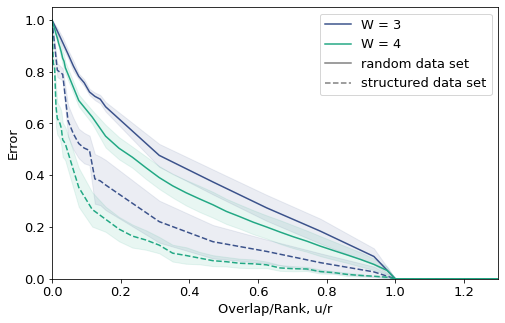}
\caption{\textbf{Comparing completion error.}}
\label{fig:techmax-error} 
\end{subfigure}
\caption{\textbf{Analysing the performance of quantum kernel matrix extension using a structured data set.} For a real-world, industrially-relevant data set (see main text), the analysis performed in Section \ref{sec:completion-results-uniform} is replicated for circuit widths of 3 \& 4 on a noise-free simulator. \emph{Note:} The 13 features of the structured data set are truncated to match the number of PQC parameters for each of the 19 templates; circuits where the number of parameters exceed 13 (i.e., PQC IDs 5 to 8 for widths 3 \& 4), are excluded from the analysis.\\\textbf{Left:} The structure in the data set does not appear to materially change the rank of the quantum kernel matrices (with respect to the uniformly-random data set used in Section \ref{sec:completion-results-uniform}). Data is not aggregated over the PQCs used. \textbf{Right:} As with the random data set, completion error increases once the overlap is less than the rank, but the error is significantly lower. Data \emph{is} aggregated over the PQCs used; lines indicate the median and the shading, one standard deviation above and below.}
\vspace{-1.5\baselineskip}
\label{fig:techmax}
\end{figure}

The results of this simulation are presented in Figure \ref{fig:techmax}. Interestingly, we find that the rank of the quantum kernel matrices generated using this data set matches those generated using the synthetic data set (Figure \ref{fig:techmax-rank}). This suggests that the scaling of the matrix rank with respect to circuit width does not strongly depend on the underlying data set being encoded.

The behaviour of the completion error is also equally interesting (Figure \ref{fig:techmax-error}). Firstly, we find that once the overlap $u$ exceeds the rank of the extended matrix, completion error goes to zero. Note that this doesn't mean that the data set is without noise but that the (potentially noisy) ground truth matrices are able to be replicated exactly.
Secondly, even when the overlap is less than the rank, the completion error incurred on the structured data set is less than that on the uniformly-random one. This finding is particularly important, as it suggests that kernel matrix completion using structured data might actually still be feasible, even if the overlap fails to satisfy the necessary condition of the algorithm to ensure that perfect completion is even possible. Therefore, even for matrices with very high rank, data transfer savings (i.e., lower overlaps) are acceptable, provided some completion error is tolerable. A related interesting question is whether the particular ML model being used would tolerate this error, and the degree to which the model's inferences would remain stable (or even useful!) if completion error were allowable.

\section{Conclusion}
\label{sec:conclusions}
The intersection of quantum computing and machine learning is a fast-moving and rapidly growing research area. In recent years, researchers have investigated what advantages -- if any -- exist by giving classical machine learning algorithms access to ``quantum kernels" (similarity measures calculated using quantum computers). In this work, we examined the feasibility of using quantum kernels for processing streaming data. In particular, recognising that the amount of data which needs to be transferred to and from the quantum computer may induce latencies which are unacceptable with regards to the machine learning workflow, rendering the model unusable. For this reason, finding a way to use quantum kernels to process streaming data -- specifically, extending quantum kernel matrices to encompass new data -- is necessary.

Approximating a kernel matrix to save on computational effort is not a completely new concept: many classical kernel methods don’t scale well and employ approaches similar to matrix completion when the data set starts to get above a few thousand data points \cite{NIPS2000_19de10ad, NIPS2007_013a006f,kernelapprox2010}. 
In this work, we have showed that classical computational techniques can be used for extending quantum kernel matrices, by introducing the idea of using classical matrix completion to do this extension in a way which is entirely ``offline" with respect to the quantum computer. This means that under certain conditions only a (comparatively) small amount of data needs to be transferred back and forth.

Extending matrices for streaming data naturally induces a block-diagonal sparsity pattern; one block consisting of the original kernel matrix for the previously-seen data, and another block consisting of all the kernel entries for the new data. We showed that a completion algorithm based on a walk over chordal graphs can be used to fill in the unknown entries. The maximum-determinant completion exploits the chordality of the block-diagonal pattern, resulting in a very efficient algorithm. The overlap between the two blocks -- which, operationally, is the amount of old data which needs to be transferred -- emerges as the key quantity governing the accuracy of the completion. It is known that by necessity for \emph{any} method to recover a matrix with full accuracy, the overlap must exceed the rank. This work suggests that this necessary condition is also sufficient for the completion of quantum kernel matrices, yielding the lowest possible sampling complexity with respect to the theory. 

We empirically studied the behaviour of completion error with respect to the block overlap in the regime where the overlap is too small, and showed that the completion error increases as the overlap decreases. Further, in the presence of shot (finite-sampling) noise, our results show a degradation in completion error even with sufficiently-high overlap between the blocks. These results indicate that quantum kernel matrices \emph{can} be extended to process streaming data, but that a high number of shots is needed to ensure good completion accuracy. Whether matrix completion is robust to \emph{device} noise, from quantum kernels run on (realistically noisy) quantum hardware, is still an open research question.

In order to know \emph{a priori} what overlap is sufficient, we studied the behaviour of the rank of quantum kernel matrices. In general, the width of the circuit being used plays a crucial role in determining the rank. For Haar-random circuits, our numerical results indicate a universal scaling behaviour of the rank. We conjecture, based on the numerical evidence presented in this work, that for Haar-random circuits on $w$ qubits, the rank of $N\times N$ quantum kernel matrices generated by them is $\min(N, 4^{w})$.

For PQCs, the situation is slightly different. For them, the scaling is not exactly as conjectured for Haar-random circuits, but it is surprisingly close. We found a weak relationship between the expressibility of a PQC (a measure of how well it approximates a Haar-random circuit) and whether the rank of the quantum kernel matrices it generates is the same as for Haar-random circuits. That is, the more expressible a PQC, the more there is agreement in the rank of quantum kernel matrices.

These results suggest that the rank of quantum kernel matrices may scale \emph{exponentially} with circuit width. This poses an obstacle for using the matrix completion algorithms presented here, as the amount of data which needs to be transferred -- i.e., the necessary block overlap to ensure accurate completion -- then grows exponentially as well. 

However, for a structured, industrially-relevant data set, we find that completion error is generally lower (as compared to a uniformly-random data set) in the regime where the overlap is too low by just a little.
This suggests we may want to work in such a ``just-low-enough" overlap regime, where perfect completion will not be possible, but the approximation error is still low. This suggests that, in practice, the overlap may not need to scale exponentially with circuit width, and with a concomitant, albeit acceptable, increase in completion error. An interesting question these results raise is how the rank of quantum kernel matrices behaves on highly \emph{correlated} data, and the degree to which correlation in the data (a) lowers the rank of quantum kernel matrices, and (b) facilitates more accurate completion.

In total, this work shows that quantum kernels can be used to process streaming data, which can be viewed as extending a quantum kernel matrix using classical matrix completion. The rank of the extended quantum kernel matrix plays a crucial role in determining whether accurate completion is possible. In the regime where the amount of block overlap is less than the rank, completion error increases. Finite-sampling noise also causes completion error to increase, even when the overlap exceeds the rank. For quantum kernel matrices arising from the action of Haar-random circuits, we conjecture a universal scaling of their rank, and show that more expressible PQCs have a similar scaling.
This scaling suggests the rank of quantum kernel matrices is exponential with respect to circuit width. For a structured, industrially-relevant data set, we find that completion error is generally lower than on a random, unstructured data set when the overlap is less than the rank. This provides hope that if the overlap was made to scale polynomially with the circuit width, an acceptable amount of completion error would be incurred for performing quantum kernel matrix extension on real data.

\section{Acknowledgements}
We acknowledge the use of IBM Quantum Services for this work. We also acknowledge the Woodside Technology and Data Science teams for providing the computing resources, data and time to undertake this study.

TLS acknowledges useful conversations with Kristan Temme, Jennifer R. Glick, Bryce Fuller, and Guglielmo Mazzola (all of IBM Quantum) on this work and earlier drafts of this manuscript.

\printbibliography

\end{document}